# Delightful Companions: Supporting Well-Being Through Design Delight

**Omar Sosa-Tzec**
University of Michigan
Stamps School of Art and Design
Ann Arbor, MI 48109, USA
omarsosa@umich.edu

**Gowri Balasubramaniam**
University of Michigan
Stamps School of Art and Design
Ann Arbor, MI 48109, USA
gowrib@umich.edu

**Sylvia Sinsabaugh**
University of Michigan
Stamps School of Art and Design
Ann Arbor, MI 48109, USA
sylvs@umich.edu

**Evan Sobetski**
University of Michigan
College of Engineering
Ann Arbor, MI 48109, USA
sobetski@umich.edu

**Rogerio Pinto**
University of Michigan
School of Social Work
Ann Arbor, MI 48109, USA
ropinto@umich.edu

**Shervin Assari**
Charles R. Drew University of Medicine and Science
College of Medicine
Los Angeles, CA 90059, USA
shervinassari@cdrewu.edu

## Abstract
This paper presents three design products referred to as *delightful companions* that are intended to help people engage in well-being practices. It also introduces the approach utilized to guide the design decisions during their creation. *Design delight* is the name of this approach, which comprises six experiential qualities that are regarded as antecedents of delight. The objective of this paper is to introduce the approach and the companions, and state the two paths that have defined the future steps of this research.

## CSS Concepts
• Human-centered computing~Interaction design~Interaction design process and methods~User interface design

## Introduction
Chronic conditions have become a major public crisis in the USA [3,5]. About 90% of the health care costs in the USA relate to chronic conditions [11]. Moreover, the majority of people have more than one chronic condition, being HIV one of them. One significant problem is that many patients with chronic conditions have a low or no income and lack of an appropriate computational and information literacy [1]. The health inequity that these people suffer has a remarkable impact in their motivation and subjective well-being [7]. This situation has become the motivation for this research on how design products—label used here in the broad sense—of the everyday life support these people and their well-being. However, this research has a particular angle. This research investigates the connection between delight and these products constitution. The initial stage of this research focused on the application of a design approach referred to as *design delight.* This approach is the result of drawing on knowledge about delight from marketing, philosophy, design, and HCI, and applying interaction criticism [2] for seven years on different design products. This research applied design delight to produce three *delightful companions* to support well-being. This paper introduces design delight and the companions, and ends with a description of the next steps in this research.

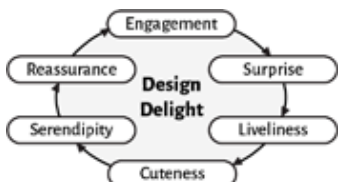

Figure 1: The six experiential qualities of *design delight* presented in the form a *canonical sequence*.

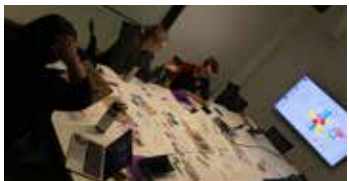

Figure 2: Designers critiquing product ideas presented as sketches.

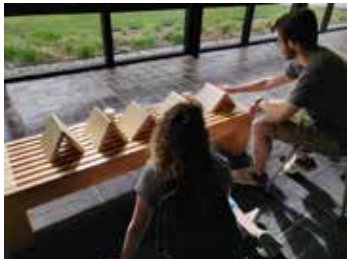

Figure 3: Designers exploring. Different materials and angles for the body of *Hey-U*.

## Design Delight
Design delight is an approach to designing delightful products and experiences. Design delight comprises six experiential qualities [8]: *engagement, surprise, liveliness, cuteness, serendipity,* and *reassurance.* Design delight broadly defines the user experience as a sequence of moments that are mostly related to one or a series of actions that the user aims to perform in order to achieve a goal or fulfill a need. In any of these moments, the user can end up feeling high pleasure and arousal, that is *delight* [9,10], as a result of experiencing one or a combination of these six experiential qualities. Design delight refers to the hypothetical moment in which engagement, surprise, liveliness, cuteness, serendipity, and reassurance occur in this order as *the canonical sequence.* (Fig. 1). Although such a sequence is unlikely to occur in actuality, design delight presents it as a conceptual lens for the user to envision how to shape the features of a product with aims to create a remarkable moment of the user experience due to the delight that it would provoke in the user.

***Engagement*** happens when the product gets to be at the center of the user's attention, sending other stimuli from the environment to the periphery. The user is captivated by the product because she can quickly identify and comprehend the features that would allow her progress towards the action that leads to the goal of interest. If this condition remains, the user remains engaged in interactions with the product, and eventually reaches a state of *flow* [6]. Once engaged, ***surprise*** happens. The product introduces an unexpected feature that appears or behaves different from other similar features found in other products of the same design genre, or that is completely new or novel to the user. The quality of surprise aims to cause a positive disruption, meaning that although it might disrupt the user's flow, it redirects the user's attention towards the product. After the user has been engaged or surprised, the features of the product may make it look like something alive, whether the features appear vibrant or show indication that the product is cognizant and has agency. These characteristics define the quality of ***liveliness***.

The product shows ***cuteness*** when it appears vulnerable, helpless, and somehow naïve to the user, mostly by alluding to an infant look and behavior. Cuteness urges the user to take care of it, to protect it, expecting that the user accept the affordances and constraints defined by the design of the product. ***Serendipity*** hap-pens when the user not only encounters an unexpected feature in the sense defined above but also discovers that this feature is convenient or beneficial for achieving the goal of interest. This quality is about making the user note the relevance of features that she had not considered before as these features are unusual or absent in the design of other products of the same genre. **Reassurance** becomes present when the product alleviates the user's uncertainty or anxiety. The user realizes how her previous actions has taken to that moment. Moreover, the user makes sense of the current appearance and behavior of the product, having no doubt about whether she has proceeded correctly regarding the goal of interest. Further, the user can foresee the next steps to perform if the goal has not been achieved yet.

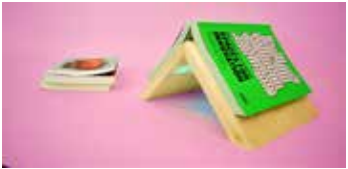

Figure 4: *Hey-U* in. its default position as a bookshelf.

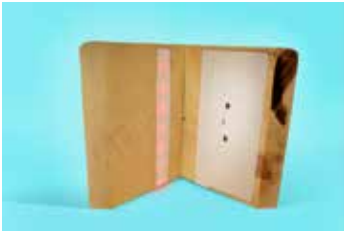

Figure 5: *Hey-U* positioned vertically so that the user can communicate with the bookshelf .

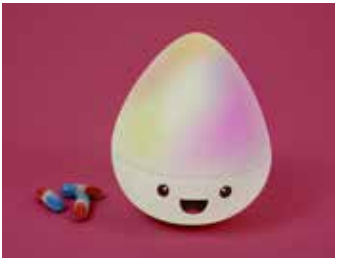

Figure 6: *Pilly-Eggy* cheering the user who has taken her medicine.

## Delightful Companions

*Delightful companions for well-being* is the label given to products intended to support people engage in practices that can help them live a happy and flourishing live. In this context, *design delight* has been explored as a generative approach. *Hey-U, Pilly-Eggy,* and the *Cocoa-Cheer Cookie Jar* are the first outcomes of this exploration (Figs. 4-9). The creation of these companions relates to a research-through-design project that seeks to support people having chronic-conditions and co-morbidity. However, as *design delight* and the research team had been recently formulated and assembled, respectively, it was decided to focus the first stage of this project on investigating how design delight supports design activities, particularly, ideation, sketching, critique, and formal and material explorations.

When these three companions were produced, the research team comprised the principal investigator (PI), two senior researchers from social work and medicine, respectively, and a group of students, all of them having a design background but from different majors, including art and design, mechanical engineering, and textile design. The PI has been involved in design practice and research for more than two decades. The two senior investigators have a solid trajectory conducting community-based research. When the project started, the PI introduced design delight through examples based on existing interactive systems. The PI also presented several sketches of delightful companions. Later, the designers in the team engaged in ideation, sketching, and critique. Each designer came up with some ideas that were critiqued by the other designers, and based on the takeaways of the critique, the designers went back to sketching (Fig. 3). This process was repeated three times. The designers utilized *Butler's six domains of self-care* [4] to situate their design decisions and maintain certain focus, particularly, because no primary research was conducted before this ideation phase. The last critique session included the two senior researchers who gave their input as experts in chronic conditions and co-morbidity. Later, the designers engaged in an iterative formal and material exploration. The designers discussed and critique the outcomes of this exploration as they were produced (Fig. 3). The three resulting companions are design mockups. The designers created a photographic storyboard to demonstrate their use in context.

*Hey-U* is an interactive bookshelf that promotes *long-distance reading relationships* and reading habits (Fig. 4). The default position of Hey-U is to put a book on it. Two linked Hey-Us communicate when one of the readers takes the book to read. Light patterns communicate the reading status on each end. Hey-U also uses its lights to urge its user to read. An asynchronous communication and amicable competition unfold as a result. Hey-U also acts like a reading companion if the user has no partner. Hey-U changes to this mode when the user changes its position to vertical (Fig. 5). The main goal of Hey-U is to create a feeling of belongingness in the user and help her improve her subjective well-being by developing a reading habit.

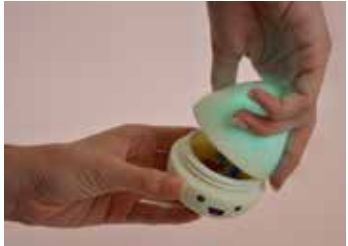

Figure 7: The user opening *Pilly-Eggy* to take her medicine.

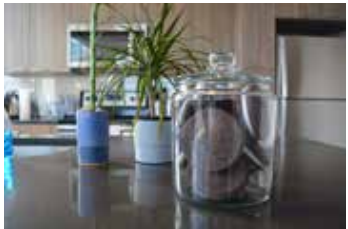

Figure 8: The *Cocoa-Cheer Cookies. Jar* left by the user in the kitchen.

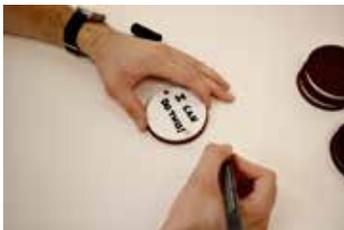

Figure 9: The user leaving a motivational message expecting to come across it again serendipitously.

*Pilly-Eggy* is an interactive pill container intended to support a person recently diagnosed with HIV and that needs to start a new medication regimen (Fig. 6). The user is expected to have Pilly-Eggy in a private space such as the bedroom or bathroom. The goal of Pilly-Eggy is to develop an affective relationship with the user. By having a cute face and a series of colorful light patterns, Pilly-Eggy seeks to celebrate the adherence to the new medication regimen (Fig. 7). As a friend who cares, Pilly-Eggy also shows discontent when the user neglects taking her medication.

The *Cocoa-Cheer Cookie Jar* contains oversized chocolate sandwich cookies stuffed with *vanilla cream* (Fig. 8). After the user has drawn one of the cookies from the jar and broken it in half, she discovers a particular type of content displayed on the surface of the vanilla cream: a photograph of the user's beloved ones, a motivational quote from other people, or a cute face for the user vent his negative thoughts and feelings. Moreover, some cocoa-cheer cookies come blank for the user to write down something motivational or significant to keep in mind and serendipitously come across later (Fig. 9).

## Next Steps

The formal and material exploration continues as these mockups move closer to be prototypes ready for evaluation. The shape of Hey-U has slightly changed as a result of integrating an Ardinuo board, a gyroscope, and rack of LED lights. Pilly-Eggy still requires finding a way to fit an arrange of LED lights that fit the cap. It is also necessary to produce a version made of medical grad plastic. There is also the plan of creating a family of Pilly-Eggies to make the pill containers works as a support team for the user. Ideally, the Pilly-Eggy of the user would communicate among each other.

Concerning the focus on chronic conditions and co-morbidity, the delightful companions are used to elicit information about how design products participate in the life of a seropositive person. When needed, the delightful companions are used as examples to initiate a conversation and co-ideation with seropositive persons. The objective is to come up with a reproducible companion that actually fits the everyday of seropositive persons.

Therefore, the development of the companions continue in order to explore, reflect upon, and illustrate *design delight.* However, it is plan to evaluate a refined version of the companions in order to gain a better understand of how delightful products support well-being practices. On the other hand, the use of this iteration of the companions to elicit information from seropositive persons is expected to lead to a new companion that would be deployed and evaluated in the field.

## References

1. Stacy C. Bailey, Rachel O'Conor, Elizabeth A. Bojarski, et al. 2015. Literacy disparities in patient access